\documentclass[manuscript]{acmart}

\AtBeginDocument{%
  }

\setcopyright{acmlicensed}
\copyrightyear{2024}
\acmYear{2024}
\acmDOI{}

\acmConference[XAIxArts 2024]{Explainable AI for the Arts Workshop 2024}{June 23, 2024}{Chicago, IL, United States}
\begin{document}

\title{Reducing Barriers to the Use of Marginalised Music Genres in AI}


\author{Nick Bryan-Kinns}
\affiliation{%
  \institution{University of the Arts London}
  \streetaddress{272 High Holborn}
  \city{London}
  \country{United Kingdom}
  \postcode{WC1V 7EY}}
\email{n.bryankinns@arts.ac.uk}
\orcid{0000-0002-1382-2914}

\author{Zijin Li}
\affiliation{%
  \institution{Central Conservatory of Music}
  \city{Beijing}
  \country{China}
  \postcode{100032}}
\email{lzijin@ccom.edu.cn }
\orcid{0009-0000-2052-3472}

\renewcommand{\shortauthors}{Bryan-Kinns and Li}

\begin{abstract}
AI systems for high quality music generation typically rely on extremely large musical datasets to train the AI models. This creates barriers to generating music beyond the genres represented in dominant datasets such as Western Classical music or pop music. We undertook a 4 month international research project summarised in this paper to explore the eXplainable AI (XAI) challenges and opportunities associated with reducing barriers to using marginalised genres of music with AI models. XAI opportunities identified included topics of improving transparency and control of AI models, explaining the ethics and bias of AI models, fine tuning large models with small datasets to reduce bias, and explaining style-transfer opportunities with AI models. Participants in the research emphasised that whilst it is hard to work with small datasets such as marginalised music and AI, such approaches strengthen cultural representation of underrepresented cultures and contribute to addressing issues of bias of deep learning models. We are now building on this project to bring together a global International Responsible AI Music community and invite people to join our network.

%\nbk{add the keywords/ CCS etc.}

\end{abstract}


\begin{teaserfigure}
  \includegraphics[width=\textwidth]{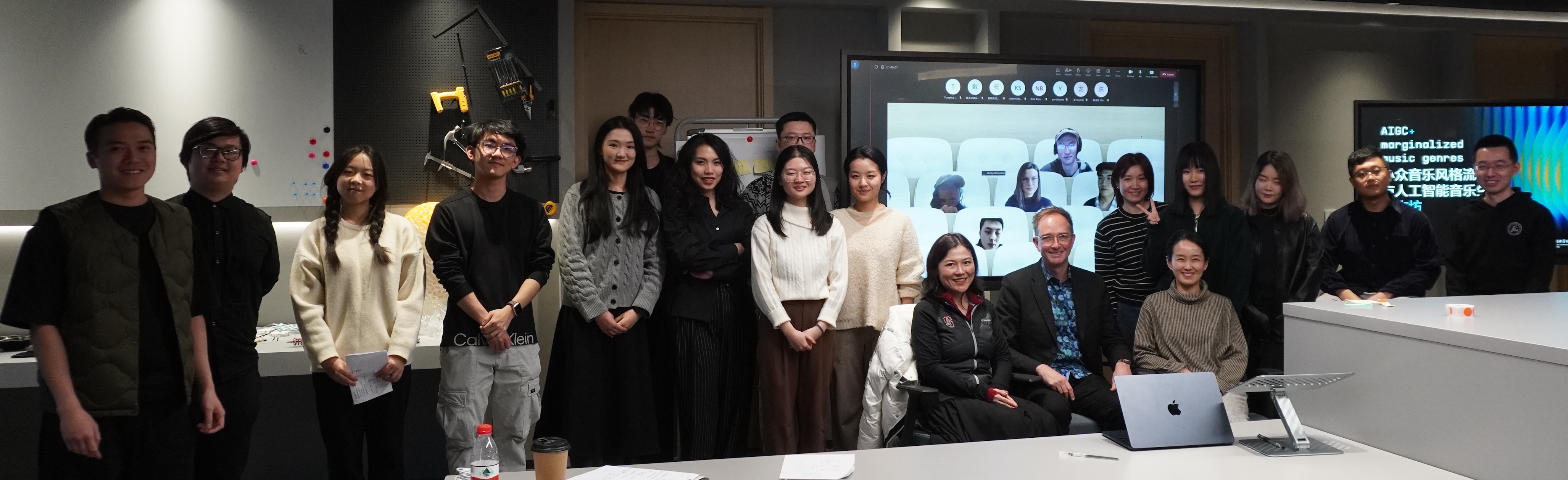}
  \caption{Workshop participants at the hybrid workshop, Tsinghua University, Beijing, China. January 2024.}
  \Description{A photo of all workshop participants together for a group photo}
  \label{fig:workshopparticipants}
\end{teaserfigure}


\maketitle

\section{Introduction}
Music is a fundamental form of human activity and artistic endeavour. The impact of AI is increasingly felt across music making from creation and production,  protection,  distribution,  to consumption. AI systems for high quality music generation and production exist in research labs and online, e.g. Magenta\footnote{\url{https://magenta.tensorflow.org}} powered by Tensorflow (Google) and Suno AI\footnote{\url{https://suno-ai.org}}. However, these systems typically rely on extremely large musical datasets to train the AI models which biases the models to genres captured in large datasets and creates barriers to generating music beyond the dominant musical genre datasets such as Western Classical music or pop music \cite{BryanKinns2020OnDP, BryanKinns2023}. For example, it is not possible to train these models on many minority culture genres such as Qin genre in China nor contemporary subcultures such as glitch or algorithmic music as the datasets are simply too small or non-existent. Whilst recent advances in AI research have started to explore low-resource AI approaches \cite{Hung2022LowResourceMG, PelinskiEmbeddedAF} which have the potential to make small datasets usable in deep learning generative models they are currently difficult for musicians understand and use. %However, such approaches have not been explored with musicians and datasets of marginalised music are not readily available for use by AI given the current emphasis on extremely large datasets.
The combination of these factors means that the rapid increase in the use of AI within the music ecosystem not only creates barriers to using small datasets and marginalised genres with AI but also that AI increasingly marginalises genres of music that are already underrepresented or not represented at all in large datasets of music.

To begin to address this gap this paper reports on a four month international research project\footnote{\url{https://nickbknickbk.github.io/UKCNAIMusicBarriers/}} which explored eXplainable AI (XAI) \cite{gunning_2016} and broader Human-Centred AI (HCAI) \cite{Shneiderman2022} and Responsible AI (RAI) challenges and opportunities of using small datasets with AI music generation. We take a broad view of XAI and include explanations of AI bias, trust, training, datasets, control, and transparency. In doing so the project asked how we make AI models for music more ethical, less biased, and more understandable and usable by people - musicians in our research.

The project focused on AI and music in the UK and China as the Creative Industries and Intangible Cultural Heritages of the UK and China provide a rich ecosystem of AI research along with substantial cultural heritage beyond the dominant forms typically used in current music AI research.

Our \textbf{research question} is: What are the eXplainable AI challenges and opportunities for using small datasets of music with AI?

\section{Data Collection}
To understand the XAI challenges of using small datatsets with AI music models we undertook interviews with seven international experts. % in AI and music to capture current state of the art. 
We then held a hybrid workshop between the UK and China hosted at the Institute of Data and Intelligent Innovation Design, Academy of Arts and Design, Tsinghua University, Beijing, China, in Jan 2024 (Fig \ref{fig:workshopparticipants}). Thirty-seven participants were brought together to build a community to share knowledge, experience, and practice around using AI with marginalised music. The workshop included brainstorming about challenges and opportunities, and case study sharing.
We also commissioned a new piece of music\footnote{\url{https://vimeo.com/925184799}} composed using the Chinese Musical Instrument Database and RAVE AI\footnote{RAVE 2.3 \url{https://github.com/acids-ircam/RAVE}} to foreground the practical challenges of undertaking such creative endeavours.

\section{Results}
The interviews and workshop discussions were video recorded and transcribed and then analysed using Thematic Analysis \cite{BraunClarke2006} to identify challenges and opportunities for UK-China collaboration in this area. Four themes were identified as outlined below followed by the identified challenges and opportunities.

%\nbk{need to go through and connect these issues to XAI/ RAI concerns}

\textbf{Theme 1: Access to AI-enabled Music Exploration.}
Barriers that hinder the widespread utilization of AI in music exploration with marginalised music genres included: \textbf{Ethical \& economic constraints}\label{sec:subtheme_ethical} such as ethical data use and the cost of music collection and analysis. 
Here the XAI question is how to explain where the data came from and limitations on its ethical use. 
%Challenges arise from the ethical and economic aspects of AI Music utilization. These include the dilemmas surrounding ethical data use [Ethical data use] and the dominance of big tech companies who build and control AI music technologies [Dominance]. Economic issues also play a significant role, as the cost of dataset acquisition and the financial implications of using proprietary music data and using human labor to collect and analyze data can limit the scope of AI music projects [Limited funding]. Consideration of financial and time investment and output [Economic viability] also limits the utilization of niche music data. 
\textbf{Knowledge and skills barriers}\label{sec:subtheme_knowledgebarriers}
The need for an in-depth set of multidisciplinary skills and knowledge to balance music domain knowledge with AI skills were highlighted as a critical barrier. This raises XAI questions of how to understand and control AI models.
For example, AI music scholars face challenges in fully harnessing AI's potential due to the need for specialized musical knowledge and technical expertise which is particularly evident when attempting to engage with non-mainstream or regional music genres.%, which require a deep understanding of specific musical traditions [Representation selection], cultural meaning embedded in music [Integrating cultural learning] and at the same time a the technical understanding of AI performance [Rethinking evaluation metrics] to build AI music models and research effectively. 
\textbf{Technical barriers}\label{sec:subtheme_technicalbarriers} such as the need for high-quality and large datasets for effective AI model training, and the high computational demands for AI training. This raises XAI questions about explaining and understanding training of low-resource models with smaller datasets to reduce the dependency on deep learning models.

\textbf{Theme 2: Stakeholder dynamics.}
How musicians want to interact and engage with AI models was seen as critical to their success and use from a HCAI \cite{Shneiderman2022} perspective - AI's capacity to extend and enrich musicians' creative vision requires technologists to focus on understanding and addressing musicians' needs.
In doing so we can use XAI to better explain to musicians how the AI models work, and to design user interfaces to allow for more meaningful control of AI by musicians.
%At the same time narrow research interests of AI music researchers limits the ability for researchers to share datasets and AI skills.
%\nbk{
%This theme examines the interaction between technological advancements and audience engagement, highlighting 
Participants highlighted how experimentation and critical reflection by musicians can potentially inform explanation of AI music models to make AI music generation and exploration more accessible and inclusive.%, especially for underrepresented genres.

%\subsubsection{Musician critique and self-reflection}
%AI's capacity to extend and enrich musicians' creative vision requires technologists to focus on understanding and addressing musicians' needs [Emphasizing user needs]. AI technologies offer new avenues for musical exploration and innovation [Creative influence of AI on music]. When AI interacts with diverse musical traditions, it was suggested that critical thinking could help to foster respectful and authentic representations [Cultural sensitivity]. 

%\subsubsection{Narrow research interests}\label{sec:subtheme_narrowresearch}
%A tendency was noted in researchers to focus research efforts on specific music genres or data produced by one musical instrument. This can restrict engagement with, and utilization of, a wider range of datasets.

\textbf{Theme 3: Technology innovation.}
Innovative uses of AI along with innovative AI techniques offer new opportunities for using AI with marginalised music. For example, participants noted that AI can offer \textbf{novel and innovative} opportunities for co-creation and co-creativity development. This will require careful XAI design to ensure that the interaction is intuitive and understandable for musicians.
% [Collaborative AI-music creativity] and offer opportunities for experimental music production methods and acoustic experiences for musicians and AI technologists [Experimentation with AI music].  AI can also enhance musical analysis through data processing [Enhancing musical analysis and discovery]. Researchers reported invented new methods and tools to handle underrepresented musical data [Inventing emerging AI] and have discovered ways to use small, limited resources more effectively [Using small dataset effectively]. 
%\subsubsection{Techniques for training AI on limited resources}
To address the practical challenges of \textbf{training AI on limited resources}, participants suggested fine-tuning large models with small datasets to reduce bias, augmenting small datasets using synthesised data, compressing datasets to extract the most salient musical features, or using transfer-learning to transfer learning to use features of large models with small datasets. In addition it was suggested that more \textbf{experimentation with AI model selection} would benefit music generation with small datasets rather than relying on deep learning approaches. The requires more explanations of small AI models and their potential use. 
%[Data augmentation and compression]. Extensive experimentation with AI model selection and fine tuning will be required to optimise them given the unique nature of many musical genres and datasets [Experimentation with model selection] rather than attempting to use large general purpose models. In terms of the use of low-resource AI models, strategies were identified as showing promise for small datasets such as marginalised music. These included using large models to generate broad rules for music generation which are then optimised in low-resource models [Low resource adaption] and using transfer learning to apply parts of a large model to a small dataset [Transfer Learning].

\textbf{Theme 4: Sustainable and Actionable Development in AI Music Ecosystems.} 
A number of strategies for creating a supportive environment for AI music research and development were highlighted by participants. These primarily focused on XAI aspects of academic and scholarly support as well as community and stakeholder engagement and support which itself relies on XAI to help understand the positive potential of AI music making. For example, using research outcomes as an XAI voice to advocate for ethical musical influence on culture and humanities. %Participants advocated for the ethical involvement of a wide spectrum of stakeholders beyond the academic sphere and reflected on best practices working with particular communities.

\section{Discussion and Conclusions}
%\nbk{need to connect with XAI/ RAI}

Regardless of the nature of the marginalised music genre, they all suffer from increased marginalisation by deep learning AI. In addition to the lack of existing large datasets of marginalised music genres there many features of these genres which are simply difficult to encode into forms of representation that might be used in deep learning model training such as MIDI. Moreover, many of these genres are shared, learnt, and passed from generation to generation through oral tradition rather than being written down which again makes them less amenable to use with AI models than music genres which are well described and documented by conventional musical notations and encoding.

The \textbf{Key Take Away} from our analysis is that one AI model or architecture will not be able to handle the wide variety of music styles, genres, heritage, cultural contexts, and forms of notation and documentation. This challenges the dominant contemporary discourse on the wide application and general use of deep learning models such as large language models and catalyses opportunities for collaboration. From the analysis of workshops and interviews we identified opportunities for collaboration between the UK and China on reducing barriers to the use of marginalised music with AI including: 1) New techniques for dataset building which relies on XAI to understand the training constraints of AI models; 2) New approaches to AI model training requiring XAI to explain how the training works; 3) Sharing examples of small dataset use which needs XAI to explain how the AI model is trained and used; 4) Interdisciplinary collaborations; 5) Community and stakeholder engagement.
%In this project we built the seeds of a UK-China community of researchers and musicians interested in using AI models both for generation and classification of marginalised music not currently usable in major deep-learning models. We identified and scoped a number of ways that artists could access and use datasets of marginalised music with AI from data augmentation and transfer learning to AI models such as RAVE. 
To bring together XAI resources for the use of marginalised music with AI we suggest building an online repository to include:
\begin{itemize}
    \item Descriptions of a variety of musical genres including rich descriptions of the genre, the associated musical traditions, musical notations (if any), musical practices and conventions, and the cultural context of the genres.% This could be documented using, for example, datasheets for datasets \cite{gebru2021datasheets}, to provide an open and structured way to describe musical genres and their context.
    \item Musical content and corpora for the music genres. For example, in the style the Dunya project\footnote{\url{https://dunya.compmusic.upf.edu}} which includes audio recordings and descriptions of the recordings and cultural context as well as related software tools. %for 5 marginalised genres. This approach would benefit from including audio and symbolic music, documentation and/ or recordings of performance styles, notations where available and appropriate, mechanisms to translate between different representations, and original datasets as well as preprocessed datasets prepared for use with AI models.
    \item XAI descriptions of the suitability of AI architectures, models, and settings for the genres in the repository. %Each AI approach will be a different combination of architectures, models, and settings which is hand-crafted to the musical genre, cultural context, and intended use.
    \item Explanations of AI fine-tuning and style transfer opportunities between pre-trained deep AI models and the AI models and genres in the repository. % - this would reduce training and computation requirements when using small datasets.
\end{itemize}

The \textbf{key XAI value add} of the repository will be the explanations of interconnections between the parts of the repository. For example, XAI mapping between musical content, suitable AI models, and fine-tuning and style transfer opportunities. %It is likely that some sort of academic or non-governmental organisation will be needed to build and support such as repository over an extended period of time given that it will be culturally valuable but not necessarily financially self-sustaining. 
Finally, participants in this project emphasised that whilst it is hard to work with marginalised music and AI, such approaches strengthen cultural representation of underrepresented cultures and contribute to addressing issues of bias of deep learning models. We believe that taking an XAI approach to developing low-resource models and training mechanisms will offer more opportunities for musicians to engage with and use AI in their music making practices with small datasets of marginalised music. We hope that this would help to increase representation of underrepresented genres in the AI music ecosystem. We are now building on this project to bring together a global International Responsible AI Music community\footnote{\url{http://musicrai.org}} and invite people to join our network.

\begin{acks}
Many thanks to the interviewees and workshop participants for contributing their time and insights to this project. Any many thanks to Qiong Wu, Zixuan Xu, Shuoyang Zheng, Yan Gao, Yifan Butisk, Soham Kundu, and Zhou Zhou for supporting the project.
Funded by the AHRC SEED Fellowship grant (AH/Y000722/1) and UKRI RAI UK grant (EP/Y009800/1) for the international RAI Music community development.
\end{acks}

\bibliographystyle{ACM-Reference-Format}
\bibliography{sample-base}

\end{document}